\documentclass[twocolumn,superscriptaddress,showpacs,aps,prl]{revtex4-1} 

\usepackage{amsmath,amssymb,bm}
\usepackage{graphicx,color}
\usepackage{hyperref}
\usepackage{epstopdf}
\usepackage{mathrsfs}
\usepackage{float}
\usepackage{subfigure}
\usepackage{braket}

\newcommand{\nl}{\nonumber \\}
\newcommand{\up}{\uparrow}
\newcommand{\down}{\downarrow}
\newcommand{\la}{\langle}
\newcommand{\ra}{\rangle}

\newcommand{\Fig}[1]{Fig.\,\ref{#1}}
\newcommand{\Eq}[1]{Eq.\,\eqref{#1}}

\newcommand{\be}{\begin{equation}}
\newcommand{\ee}{\end{equation}}

\begin{document}
\bibliographystyle{unsrt}
\preprint{APS/123-QED}

\title{Bias-Induced Chiral Current and Topological Blockade in Triple Triangular Quantum Dots}
\author{YuanDong Wang}
\affiliation{%
 Department of Physics, Renmin University of China, Beijing 100872, China
}
\author{ZhenGang Zhu}
\affiliation{%
School of Electronic, Electrical and Communication Engineering, University of Chinese Academy of Sciences, Beijing 100049, China
}
\author{JianHua Wei}\email{wjh@ruc.edu.cn}
\affiliation{%
 Department of Physics, Renmin University of China, Beijing 100872, China
}
\author{YiJing Yan}
\affiliation{%
Hefei national laboratory for physical sciences at the microscale,
University of science and technology of China, Hefei, Anhui 230026, China
}
\date{\today}
\begin{abstract}

We theoretically investigate the quantum transport properties of
a triangular triple quantum dot (TTQD) ring connected
with two reservoirs by means of analytical derivation
and accurate hierarchical--equations--of--motion calculation.
A bias-induced chiral current in the absence of magnetic field
is firstly demonstrated, which results from that
the coupling between spin gauge field and spin current
in the nonequilibrium TTQD induces
a scalar spin chirality that lifts
the chiral degeneracy and thus the time inversion symmetry.
The chiral current is proved to oscillate with bias
within the Coulomb blockade regime,
which opens a possibility to control the chiral spin qubit by use of purely electrical manipulations. Then, a topological blockade of the transport current due to the localization of chiral states is elucidated by spectral function analysis. Finally, as a measurable character, the magnetoelectric  susceptibility in our system is found about two orders of magnitude larger than that in a typical magnetoelectric material at low temperature.

\end{abstract}

\maketitle
%
In condensed matter physics,
the geometric (Berry) phase coherence
may strongly affect the charge dynamics.
A case in point is the anomalous Hall effect 
in ferromagnetic metals, showing a
nonzero transverse resistivity, even
there is no external magnetic field.
The geometric phase of Bloch wave functions plays
a major role in this phenomenon \cite{Tag012573}.
A nontrivial spin texture in ferromagnetic metals
produces a gauge flux that can be incorporated
into transfer integrals by additional phase factors \cite{And592,Nag902450}.
In literature, this anomalous contribution had been attributed to
the spin-orbit interaction and spin polarization of conduction electrons.
In the strong Hund--coupling limit, the conduction electron spin
aligns with the impurity spin.
The resulted fictitious magnetic field,
produced by the noncoplanar spin configuration or spin chirality \cite{Tag012573,Ohg006065},
gives rise to the Hall conductivity a topological origin.
The fictitious magnetic field has a uniform component
due to spin-orbit interactions
\cite{Ye993737,Tat03113316,Tat03076806}.

\begin{figure*}[htbp]
\centering
\subfigure{
\begin{minipage}{4cm}
\includegraphics [width=0.7in]{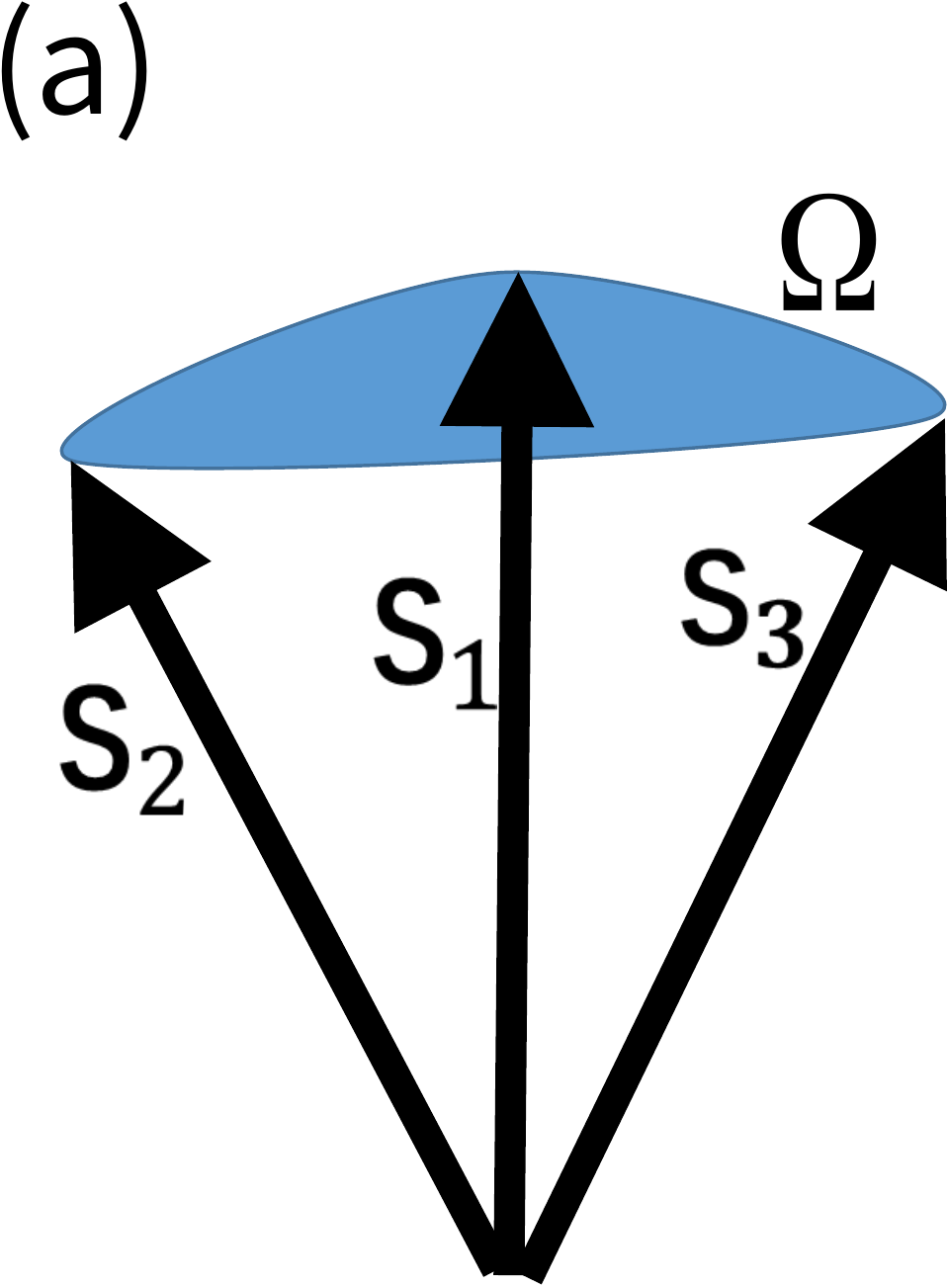}
\end{minipage}
}
\hspace{-0.4in}
\subfigure{
\begin{minipage}{5cm}
\includegraphics [width=2.0in]{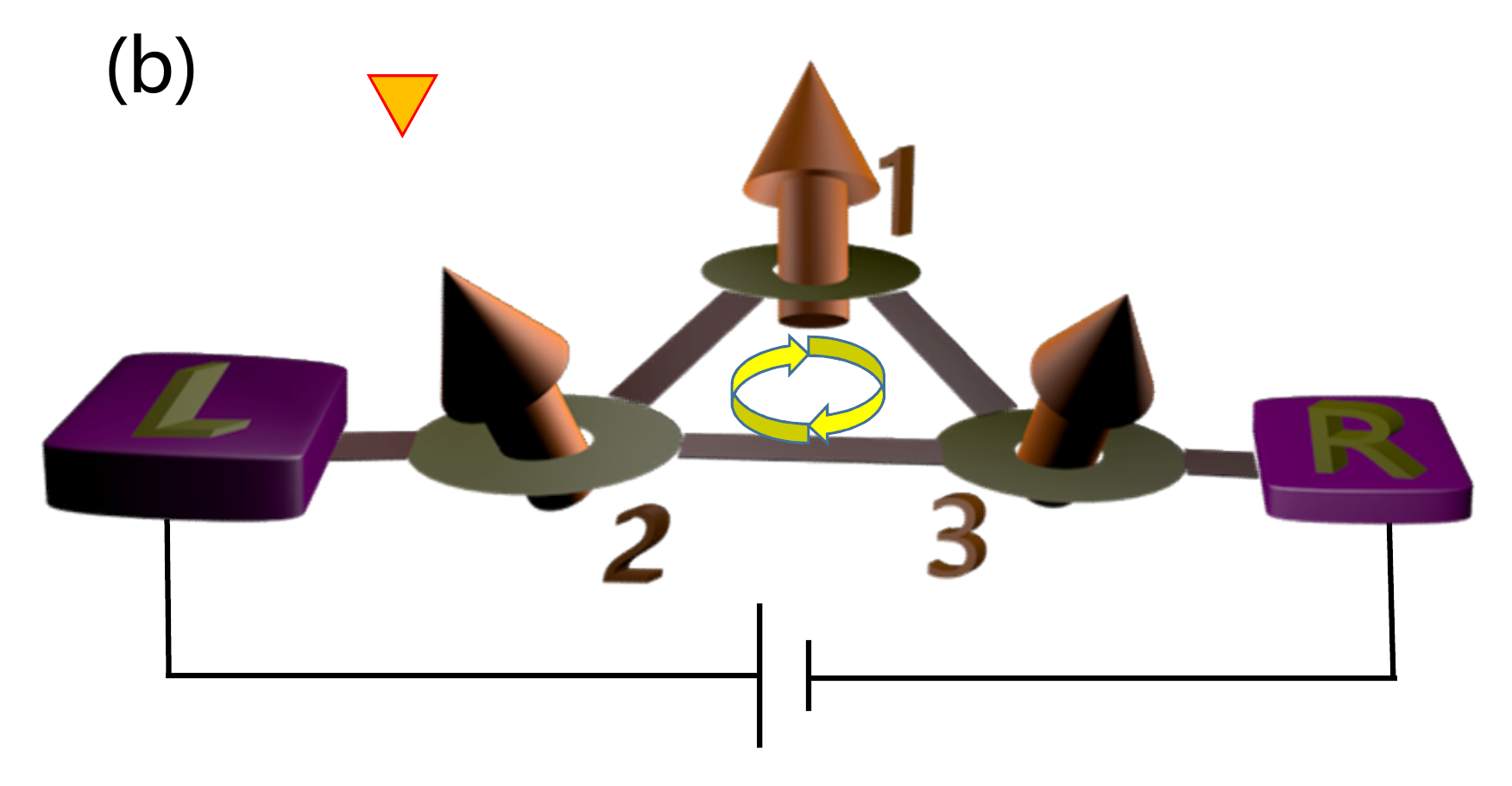}
\end{minipage}
}
\subfigure{
\begin{minipage}{5cm}
\includegraphics [width=2.0in]{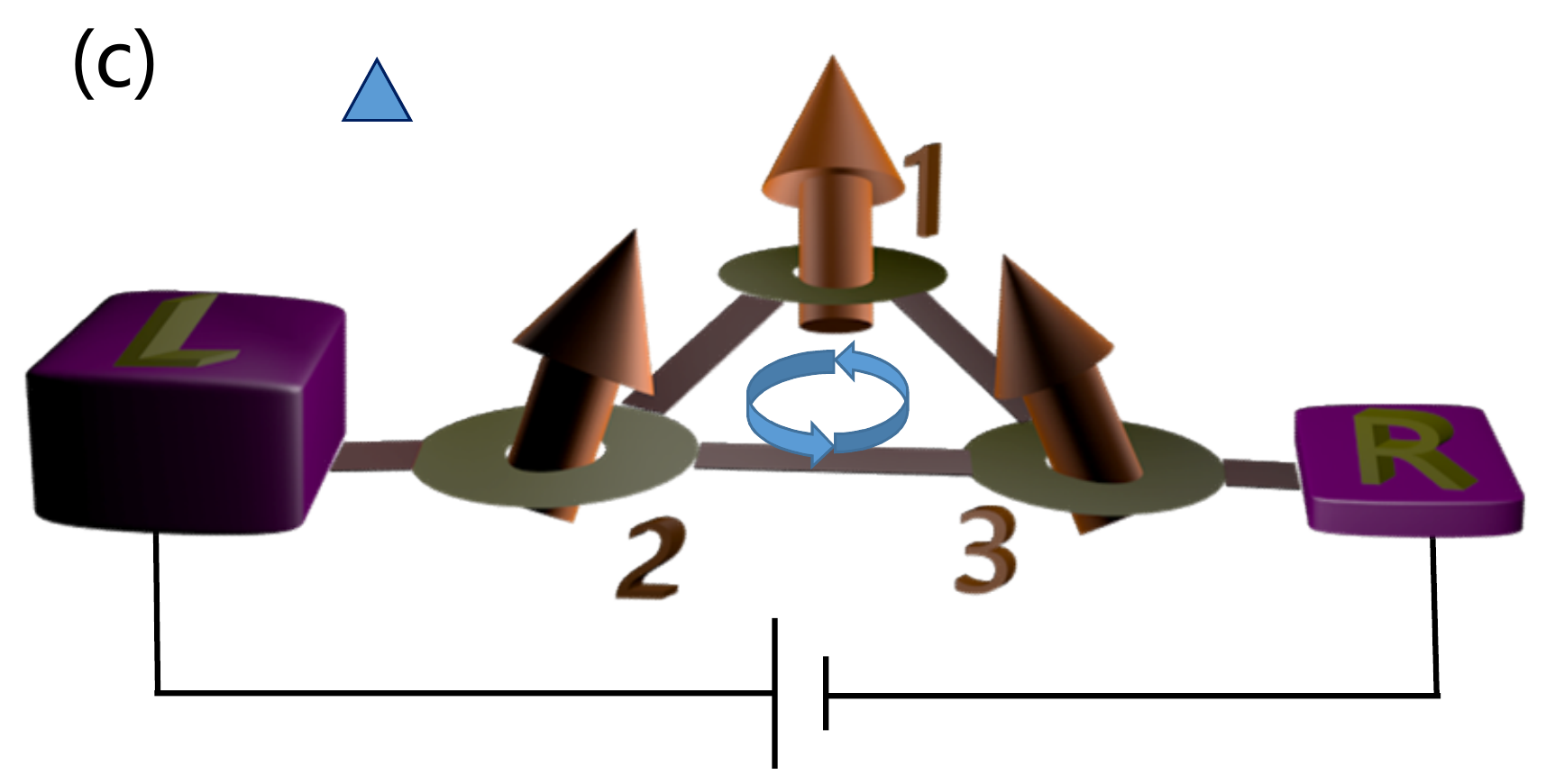}
\end{minipage}
}
\quad
\subfigure{
\hspace{-0.7in}
\begin{minipage}{4cm}
\includegraphics [width=2.3in]{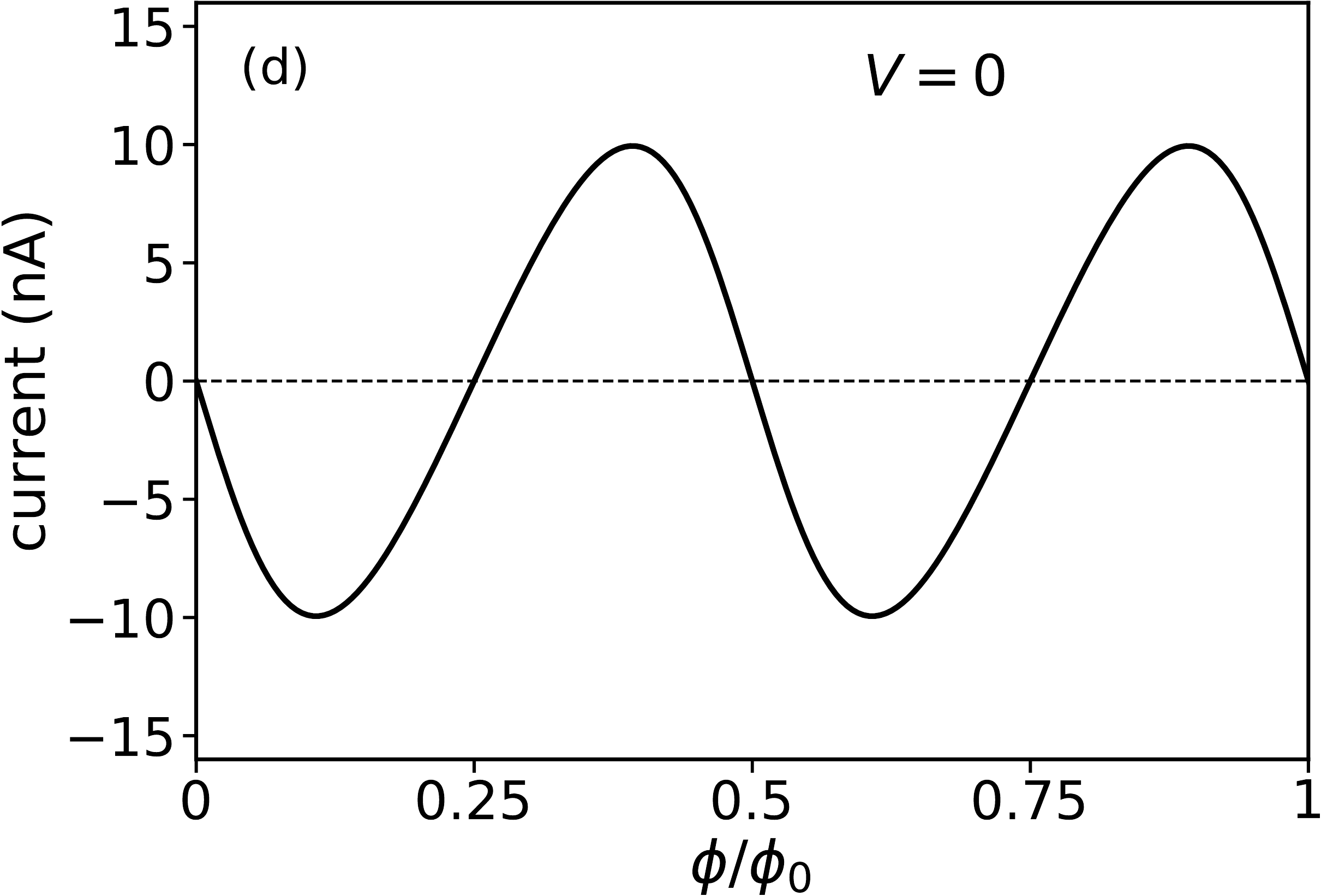}
\end{minipage}
\hspace{0.7in}
\begin{minipage}{4cm}
\includegraphics [width=2.3in]{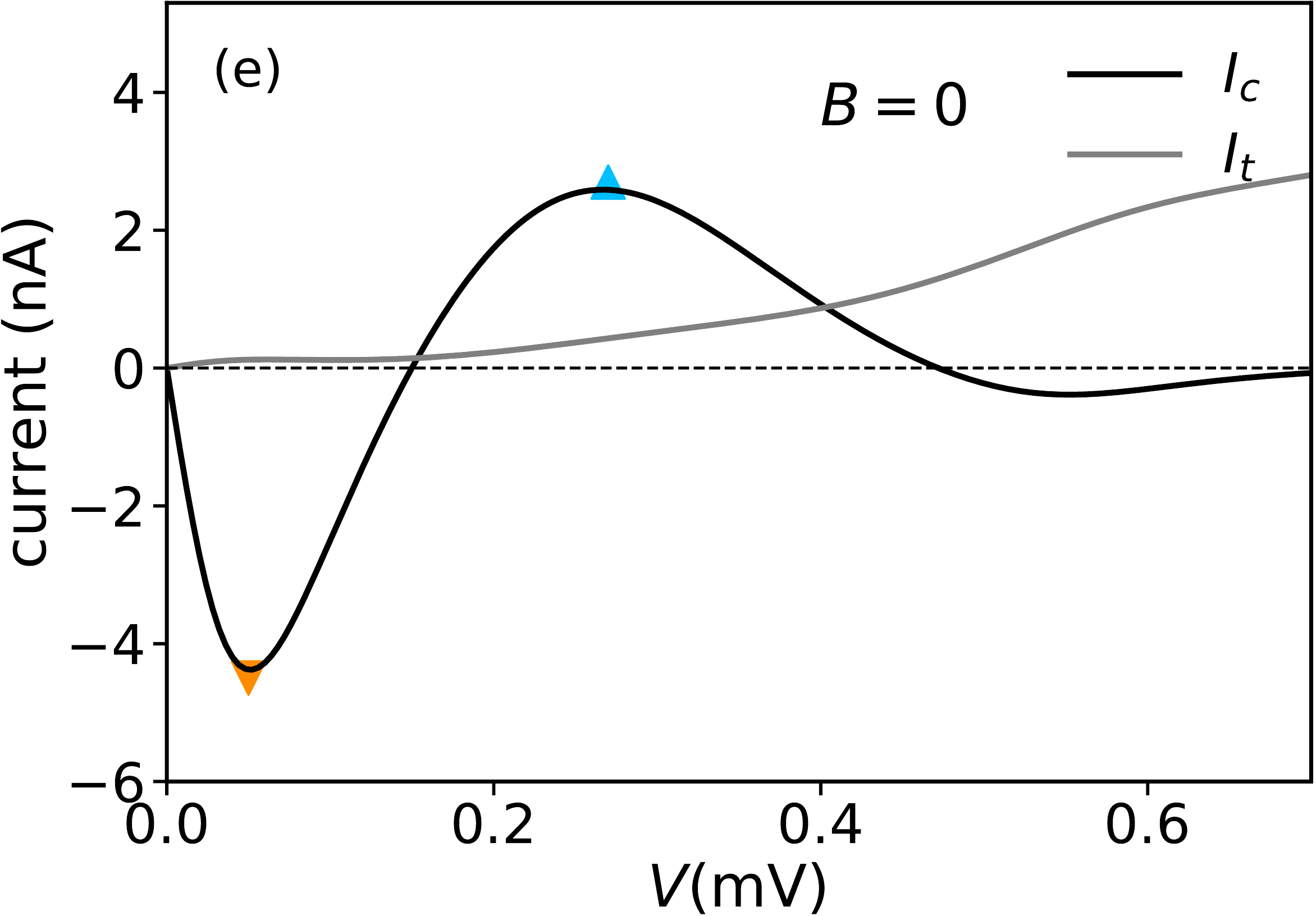}
\end{minipage}
\hspace{0.7in}
\begin{minipage}{4cm}
\includegraphics [width=2.3in]{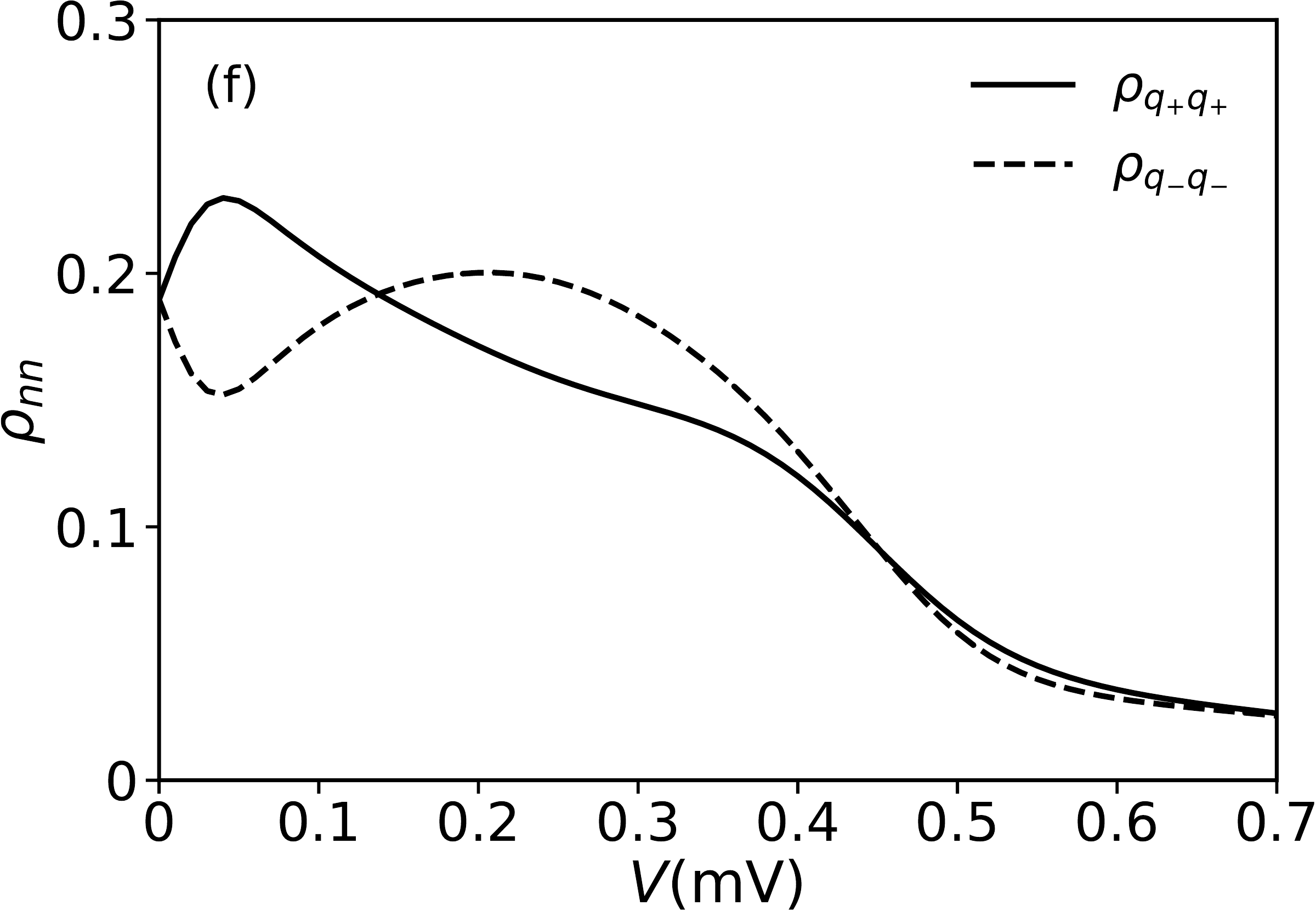}
\end{minipage}
}
\caption{(color online)
(a) Scalar spin chirality defined by the solid angle spanned by
three spins.
(b) Schematic diagram of the clockwise chiral current in the TTQD
with two dots connected to reservoirs L and R,
and (c) the anticlockwise counterpart.
(d) Chiral current as a function of magnetic flux under equilibrium condition ($V=0$). 
(e) Chiral current $I_{c}$ versus transport current $I_t$ (thin-curve),
 as functions of bias voltage $V$, without magnetic flux ($B=0$). 
(f) The $V-$dependent populations,
$\rho_{q_{+}q_{+}}$(solid curve) and 
$\rho_{q_{-}q_{-}}$(dash curve),
in the two chiral states.
The parameters are (in meV)  
$\epsilon=-0.5$, $U=1.0$,
and $t=0.25$ for the TTQD system,
$\Delta=0.025$ for the system--reservoirs coupling strength,
and $k_{B}T=0.05$.
}\label{fig1}
\end{figure*}

 Consider a minimal
chiral spin model, as shown in \Fig{fig1}(a),
with three local spins, $\bm{S}_{1}$, $\bm{S}_{2}$ and  $\bm{S}_{3}$,
whose axes are tilted away from the overall magnetization axis.
This triple triangular quantum dots (TTQDs) system
is a nonmagnetic structure.
However, when a conduction electron moves
in the background of those spins,
the phase factor picked up by this
electron is given by $e^{i\Omega/2}$,
where $\Omega$ is the solid angle subtended by
the three spins on the unit sphere.
The chiral spin state is 
closely related to
chiral-spin-liquid and superconductivity states \cite{Wen8911413}.
The geometric phase acts as the gauge field
with two essential consequences.
One is the anomalous Hall effect \cite{Tag012573}, and
another is the chiral current of
$I_{c}\propto \bm{S}_{1}(\bm{S}_{2}\times \bm{S}_{3})$.
%
We will demonstrate how to obtain chiral current
in a nonmagnetic TTQD structure
without magnetic field; see \Fig{fig1}(b)-(c).
Moreover, we uncover a novel topological blockade effect
due to this topological current.

TTQDs are composed of three coupled quantum dots
in a triangle form, with two or three reservoirs connected to them.
As the smallest artificial molecule with topological properties,
TTQDs received extensive studies, both experimental
\cite{Rog08193306,Kot16035442,Noi17084004,Hon18241115}
and theoretical \cite{Hsi10205311,Hsi12115312,Luc14165427,Luc1610,%
Hsi12114501,Wey11195302,Wrz15045407,Nik17115133}.
They are prominent candidates for research
on various quantum interference effects in the strong correlation
regime \cite{Wey11195302,Nik17115133}.
Remarkably, TTQDs have shown potential applications
in quantum computing based on the qubits
encoded in chiral spin states \cite{Gim09205311,%
Hsi12115312,Luc14165427,Luc1610}.
The chiral qubit is embedded in a decoherence-free subspace
and is immune to collective noises \cite{Vio012059},
and thus robust against random charge fluctuations \cite{Hsi10205311}.

 However, the manipulation of chiral qubit or chiral spin
state is challenging in general.
Existing proposals \cite{Hsi10205311,Hsi12115312,Luc14165427,%
Luc1610}
engage a perpendicular magnetic field
to split the degeneracy of left- and right-hand chiral states.
Nevertheless, the application of external magnetic field
would not be practical for quantum computing,
due to its incompatibility
with the large-scale integrated circuit.
Moreover, it is generally very hard to localize
the required oscillating magnetic fields
for quantum gate or qubit manipulation.

 In principle, an applied magnetic field
is not a necessity since the gauge field
from the aforementioned geometric phase could play the same role.
Motivated by this insight,
we propose the manipulation on the chiral qubit
and chiral spin state
via the bias--voltage induced
chiral current. 
We will demonstrate this proposal
with the Anderson triple-impurity model TTQDs
that are accurately evaluated and
thoroughly analyzed.


 Figure \ref{fig1}(b) or (c) depicts
our TTQD structure in a quantum transport setup.
Each QD is in the local moment regime with a $\frac{1}{2}$-spin.
The QD2 and QD3 couple to electronic reservoirs L and R,
respectively.
The total composite Hamiltonian,
$H_{\rm T}=H_{\rm{dots}}+H_{\rm{res}}+H_{\rm{coup}}$,
is described by the Anderson impurity model,
in which
\be\label{HDots}
H_{\rm{dots}}=\sum_{j,k=1}^{3}\sum_{s=\uparrow,\downarrow}t_{jk}
 \hat d_{js}^{\dagger}\hat d_{ks}
 +\sum_{j=1}^{3}U_{j}\hat n_{j\up}\hat n_{j\down}.
\ee
Here, $\hat n_{js}\equiv \hat d_{js}^{\dagger}\hat d_{js}$
is the number operator for an electron
occupying the specified on-dot spin--orbital.
For clarity, let the TTDQ hold $C_{3v}$ symmetry, with
$t_{12}=t_{23}=t_{31}=t$, whereas
$U_{j}=U$ and $\epsilon_{j}\equiv t_{jj} =-U/2$
for the on-dot Coulomb repulsion  and energy, respectively.
The non-interaction Fermion reservoir is described by $H_{\rm{res}}=\sum_{\alpha\in {\rm L,R}}
\sum_{\kappa s}(\epsilon_{\alpha \kappa s}+\mu_{\alpha})
 \hat c_{\alpha \kappa s}^{\dagger}\hat c_{\alpha \kappa s}$,
with $\mu_{\rm L}=eV/2=-\mu_{\rm R}$
The TQD system--and--reservoir coupling is described by
$H_{\rm{coup}}=\sum_{\kappa s}(t_{{\rm L}2}\hat c_{{\rm L}\kappa s}^{\dagger}\hat d_{2s}
+t_{{\rm R}3}\hat c_{{\rm R}\kappa s}^{\dagger}\hat d_{3s}+{\rm H.c.})$.

 In the following, we will first investigate the chiral
current induced by a magnetic field applied to
the isolated TTQD.
By doing that, we unambiguously
identify the chiral
current operator, $\hat I_{c}$ [cf.\ \Eq{CCut}],
which will also be used in
the bias voltage--induced chiral current evaluations.
Let us start with pristine TTQD in the absence of magnetic field and reservoirs.
The ground state of the isolated TTQD is four-fold degenerate,
with spin configuration  being 120$^{\circ}$ between
neighboring spins without chirality.
That is, the three spins are coplanar with the degenerate chiral states.
When a perpendicular magnetic field is applied,
a flux threads the ringlike TTQD structure.
A $t$-$J$-$\chi$ Hamiltonian can be derived by treating $t$ in $H_{\rm{dots}}$ perturbatively \cite{SM1, Sca05032340}:
\begin{align}\label{HEFF}
H_{\rm{eff}}&=-t(1-n)
 \sum_{jk,s}(\hat d_{js}^{\dagger}\hat d_{ks}+{\rm{H.c.}})
\nl&\quad\,
 +J\sum_{j<k} 
  (\hat{\bm{S}}_{j}\hat{\bm{S}}_{k}-\frac{1}{4}\hat n_{j}\hat n_{k})
 +\chi\hat{\bm{S}}_{1}(\hat{\bm{S}}_{2}\times \hat{\bm{S}}_{3}).
\end{align}
Here, $n$ is the average electron occupation number on each dot.
In the half-filling situation,
the $t-$term vanishes due to $n=1$.
The $J$-term is the usual Heisenberg exchange interaction,
with $J=4t^{2}/U$.
The last term is chiral, with
$\chi=24t^{3}\sin(2\pi\phi/\phi_{0})/U^{2}$,
where $\phi$ is the magnetic flux
enclosed by the TTQD,
and $\phi_{0}=hc/e$ is the unit of quantum flux.
It has been shown  that for TTQDs
the chiral operator reads \cite{Wen8911413}
\[
\hat{\bm{S}}_{1}(\hat{\bm{S}}_{2}\times \hat{\bm{S}}_{3})
=\frac{1}{2i}\!\sum\limits_{suv}\!\hat d_{1s}^{\dagger}
 (\hat d_{2s}\hat d_{2u}^{\dagger}\hat d_{3u}\hat d_{3v}^{\dagger}
  -\hat d_{3s}\hat d_{3u}^{\dagger}\hat d_{2u}\hat d_{2v}^{\dagger})
  \hat d_{1v}.
\]
It splits the four-fold degenerate ground state
in the total spin-$\frac{1}{2}$ subspace into
two chiral--states pairs.
One is the minority spin circling
clockwise ($+$) and anticlockwise ($-$) pair,
$\ket{q_{\pm}}=\frac{1}{\sqrt{3}}(\ket{\up\down\down}
+e^{\pm\frac{i2\pi}{3}}\ket{\down\up\down}
+e^{\pm\frac{i4\pi}{3}}\ket{\down\down\up})$.
Another pair for $S_{z}=-1/2$ are similar but with all spins flipped. For $S_{z}$ degeneracy denote $\ket{q_{+}^{1/2}}$ as $\ket{q_{+}}$ for simplicity and that of $S_{z}=-1/2$ are of same values.
These are the eigenstates of
the TTQD chiral operator or the last term of \Eq{HEFF}
that describes the electrons circular
transfer difference between clockwise and anticlockwise directions.
We identity the chiral current operator,
\be\label{CCut}
 \hat{I}_{c}=-\frac{24e}{\hbar}\frac{t^{3}}{U^{2}}\hat{\bm{S}}_{1} (\hat{\bm{S}}_{2}\times \hat{\bm{S}}_{3}).
\ee
It follows the Hellman-Feynman theorem for
chiral current, $I_{c}
=-\frac{e}{\hbar}\la\frac{\partial\hat H_{\rm{dots}}}{\partial{\phi}}\ra
=-\frac{e}{\hbar}\frac{\partial F_{\rm{dots}}}{\partial{\phi}}$,
with $F_{\rm{dots}}$ being the
magnetic field induced free--energy \cite{Bye6146}.
TTQD constitutes the shortest loop where each dot is in local moment region.
The coefficient $t^{3}/U^{2}$ is
the lowest-order nonvanishing contribution to the circling current.
 Figure \Fig{fig1}(d) depicts the chiral current
as a function of flux at equilibrium state.
It shows a double period with flux.
In short, the chiral operator breaks the symmetry of TTQD
from $C_{3v}$ into $C_{3}$ and induces a chiral current.

{\it Chiral current induced by bias voltage.}
We will elaborate below that
for the open TTQDs, a finite applied bias voltage
alone could also break the chiral symmetry and drive a nearly
pure chiral current. Lai {\it et al.} report that away from local magnetic moment regime, internal charge
current circulation can spontaneously emerge, with a non-monotonic behavior of transport current when the circulation reverses \cite{Lai1847002}. Here we focus on the electron transport in local moment regime, in which the phase coherence of spins plays an important role.
Let us start with the accurate numerical
results via the well--established
hierarchical equations of motion (HEOM)
approach \cite{Jin08234703,Hu11244106,Li12266403,Ye16608}.
High-order tunneling processes such as cotunneling \cite{Wrz15045407} and many-body tunneling \cite{hou172486} have been well handled by HEOM.
The present TTQD in study has $\epsilon=-U/2=-0.5$ meV.
The couplings between electrodes and TTQD are
set to $\Gamma_{\rm L}=\Gamma_{\rm R}=\Delta=0.025$ meV.
The temperature is $T=0.6$\,K,
which is far above the Kondo temperature that is
about $T_{K}\sim 3.6\times 10^{-4}$ mK.

 Figure \ref{fig1}(e) depicts the calculated chiral current
$I_c$ as a function of bias voltage $V$.
The resulted $I_c(V)$ shows oscillations
at $V<U/2$.
This is the Coulomb blockade regime for
the present TTQD in study.
Further increasing bias ($V>U/2$) will push
the system out of the Coulomb blockade regime,
where $I_c$ gradually decreases to zero and meantime
the transport current $I_t$ increases.

 Remarkably, our results show
that it is possible to control
the chiral spin qubit by use of purely
electrical manipulations
without magnetic field involved.
In the Coulomb blockade regime ($V<U/2$),
the lead--dressed ground state under bias
would be the aforementioned
chiral pairs $\{|q_{\pm}\ra\}$; see \Fig{fig1}(b) and (c).
In particular, compared to the magnetic--field
counterpart of \Fig{fig1}(d),
the bias--dressed states at $V=0.05$
and $0.25$\,mV are associated with the maximal
clockwise and anticlockwise
chiral currents, respectively.
Figure \ref{fig1}(f) reports
the reduced density matrix diagonal elements,
$\rho_{q_{+}q_{+}}$ and $\rho_{q_{-}q_{-}}$,
for the two chiral states populations as function of bias.
Evidently, the sign and magnitude
of the difference, $\rho_{q_{+}q_{+}}-\rho_{q_{-}q_{-}}$,
correlates well with the observed direction
and magnitude of chiral current
in \Fig{fig1}(e).
It is worth noting that the chiral ground states are
degenerate, $\rho_{q_{+}q_{+}}=\rho_{q_{-}q_{-}}$,
at $V=0.15$\,meV.
The resulted $I_c=0$ would have an effective
magnetic flux of $\phi^{\rm eff}=\phi_0/4$.
Physically this differs from
the scenarios of $V>U/2$, beyond the Coulomb blockade regime,
where $\phi^{\rm eff}\approx 0$ that is responsible
for the observed
$\rho_{q_{+}q_{+}}\approx \rho_{q_{-}q_{-}}$ and
$I_{c}\approx 0$.

\begin{figure*}[htb]
\centering
\includegraphics [width=2.3in]{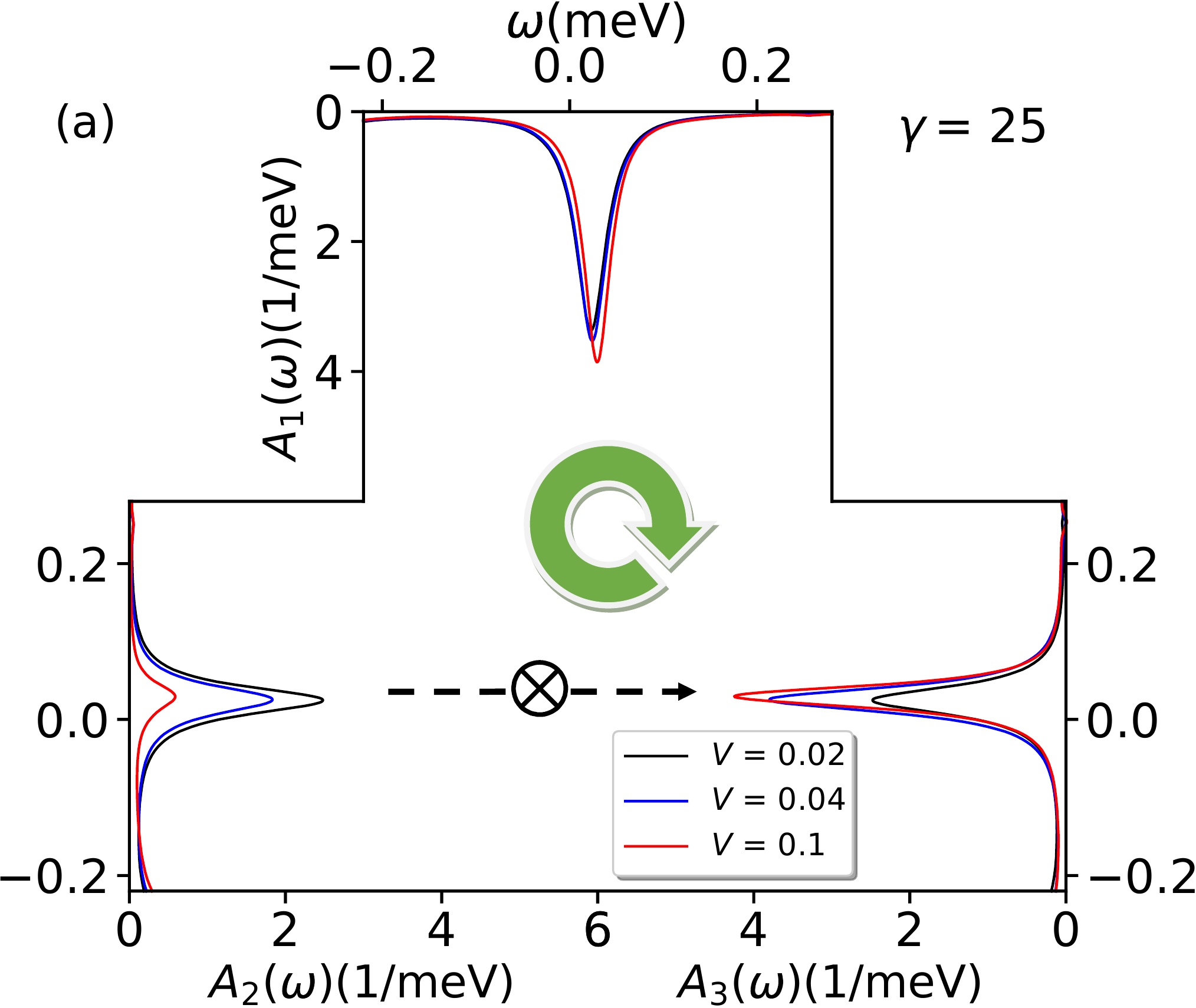}
\includegraphics [width=2.3in]{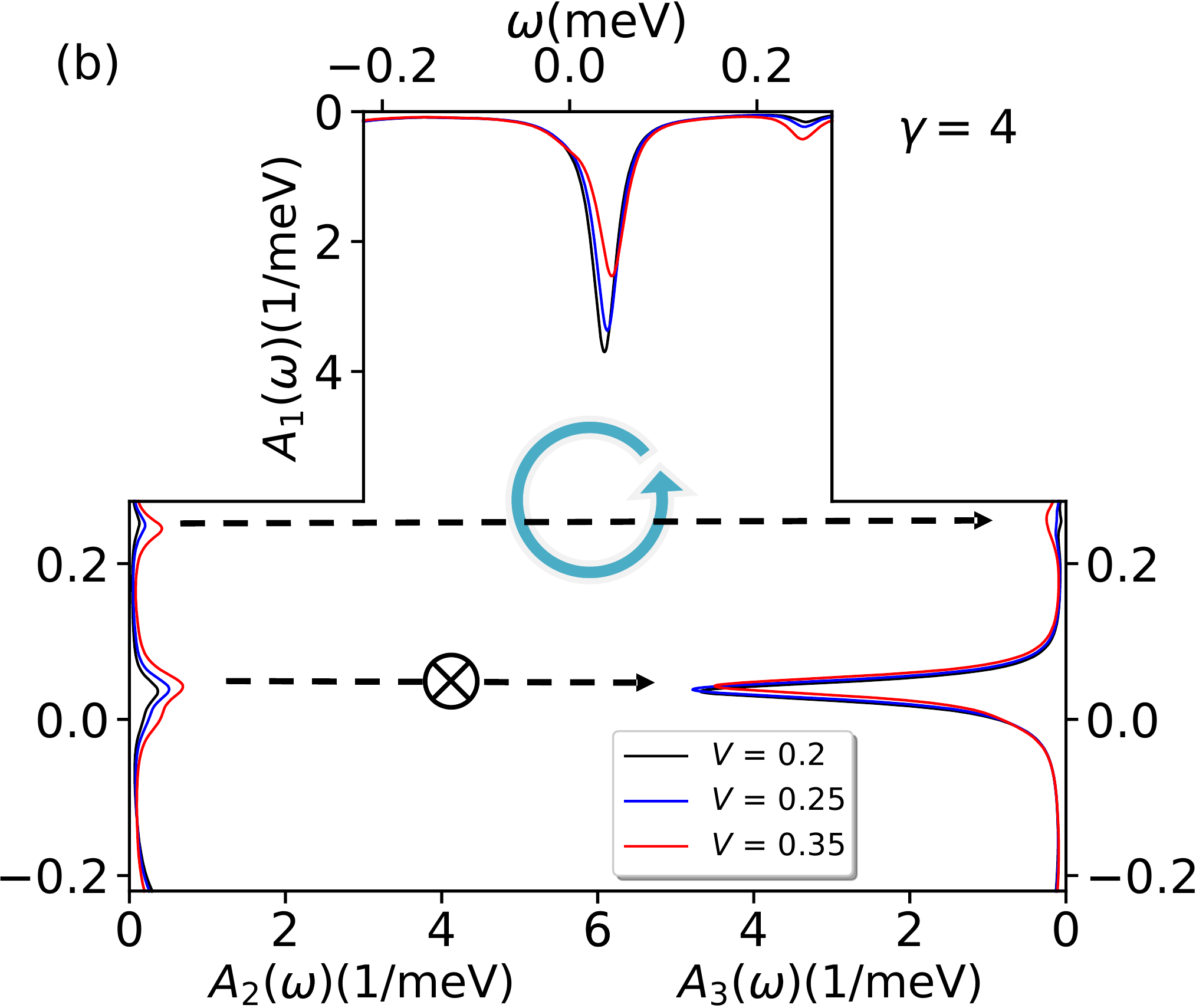}
\includegraphics [width=2.3in]{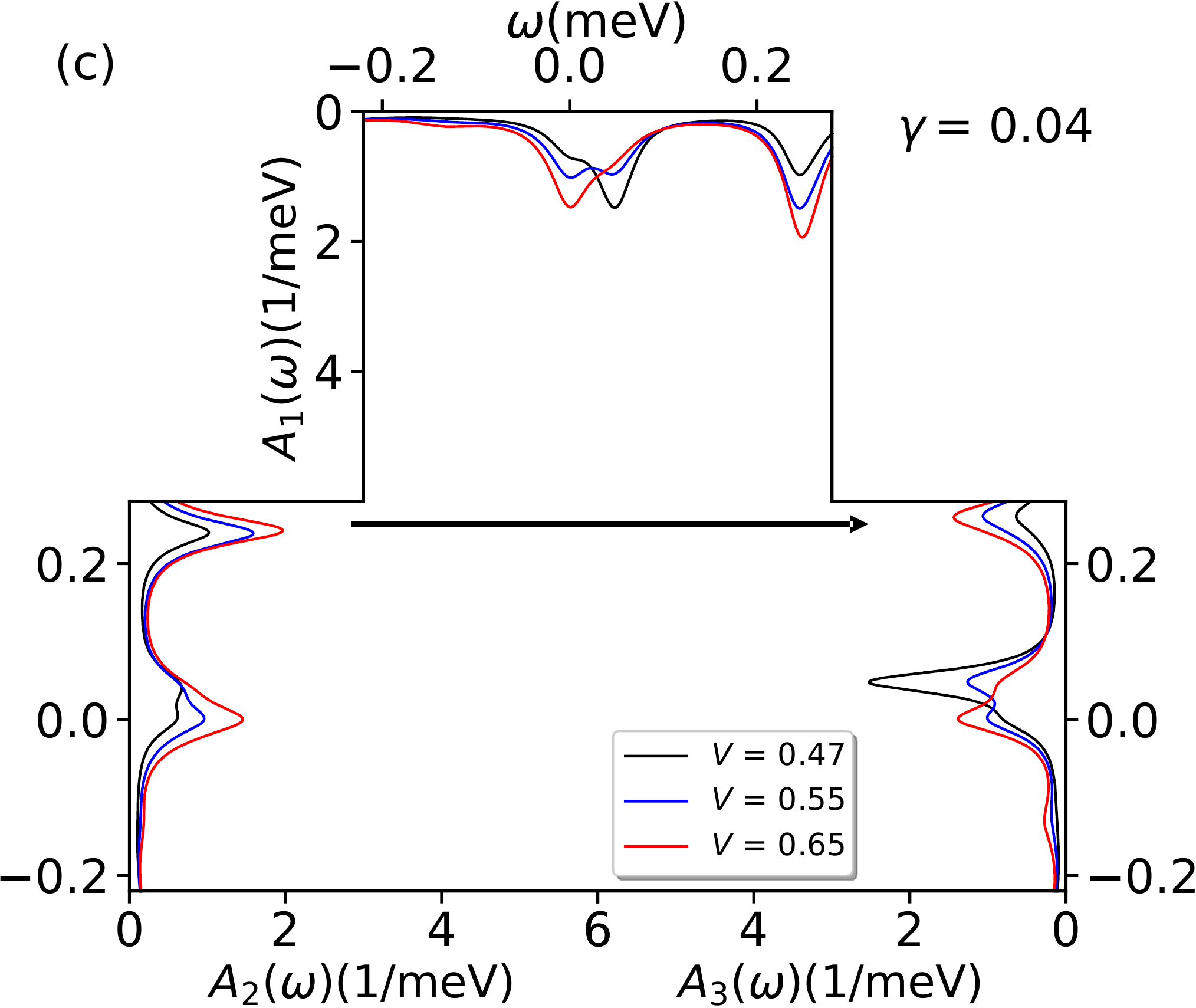}
\caption{(color online) Nonequilibrium spectral functions,
$A_{i}(\omega)=A_{is}(\omega)$, with $i=1,2,3$ and $s=\uparrow,\downarrow$,
in the (a) blockade region ($0\le V\le0.15$ mV),
(b)  coexistence region ($0.15< V\le0.4$ mV), and
(c)  conduction region ($0.4< V\le0.7$ mV).
Indicated is also the calculated value of 
$\gamma \equiv \int_\text{\tiny $V_1$}^\text{\tiny $V_2$}\! |I_{c}|dV
\big/\int_\text{\tiny $V_1$}^\text{\tiny $V_2$}\! I_{t}dV$
in each region.
}
\label{fig2}
\end{figure*}

\emph{Spin gauge field coupling analysis}.
The above observations
can be analyzed and understood as follows.
Bias breaks the inversion symmetry of the TTQD
which is required for the coupling between
spin gauge field and spin current.
The Hubbard-Stratonovich approach is adopted
to decouple the on-site Coulomb interaction
in our Anderson impurity model, as described in detail
in Supplemental Material \cite{SM2}.
To keep the spin rotation invariance,
a unitary transformation in spin space,
$\Phi_{j}=R_{j}\Psi_{j}$, is introduced.
Here, $R_{j}$ is a site- and time-dependent
SU(2) rotation matrix satisfying
$\bm{\sigma}\bm{S}_{j} =R_{j}\sigma_{z}R_{j}^{\dagger}$.
The electron operators are given in the spinor form, $\Psi_{j}^{\dagger}=(\hat d^{\dagger}_{j\uparrow}, \hat d^{\dagger}_{j\downarrow})$.
With the polar representation of the dot spin
$\bm{S}=(\sin{\theta}\cos{\phi},\sin{\theta}\sin{\phi},\cos{\theta})^{T}$,
the matrix $R=e^{\frac{i\pi}{2}}e^{-\frac{i\phi}{2}\sigma_{z}}
e^{-\frac{i\theta}{2}\sigma_{x}}e^{-\frac{i(\pi-\phi)}{2}\sigma_{z}}$
rotates the spin up state $\ket{\uparrow}$
to the direction of dot spin 
as $\ket{\bm{S}}=R\ket{\uparrow}$.
Consequently, the kinetic term of $H_{\rm{dots}}$
has a covariant form,
since $t_{jk}\Psi_{j}^{\dagger}\Psi_{k}\rightarrow
t_{jk}\Phi_{j}^{\dagger}R_{j}^{\dagger}(R_{k}-R_{j})\Phi_{k}$
and $\Psi_{j}^{\dagger}\partial_{\tau}\Psi_{j}\rightarrow
\Phi_{j}^{\dagger}(\partial_{\tau}+R_{j}\partial_{\tau}R_{j})\Phi_{j}$.
The SU(2) spin gauge field is defined through $A_{jk}\equiv -iR_{j}^{\dagger}(R_{k}-R_{j})$ and $A_{0j}\equiv R_{j}\partial_{\tau}R_{j}$.
These can be expressed in terms of the Pauli matrices as $A_{jk}=\sum_{r}A_{jk}^{r}\sigma_{r}$, where $r=x,y,z$
is the direction in spin space. The gauge field couples
to spin current through minimal coupling. The coupling energy reads
\be\label{HA}
H_{A}=\sum_{jkr}j_{jk}^{r}A_{jk}^{r}+\sum_{jr}s_{j}^{r}A_{0j}^{r},
\ee
with $s_{j}^{r}=\Phi_{j}^{\dagger}\sigma_{r}\Phi_{j}$
and $j_{jk}^{r}=-\frac{iet}{\hbar}(\Phi_{j}^{\dagger}\sigma_{r}\Phi_{k}
-\Phi_{k}^{\dagger}\sigma_{r}\Phi_{j})$
being the spin--density and spin--current operators,
coupled to $A_{0j}^{r}$ and $A_{jk}^{r}$,
the time and spacial components of spin gauge field respectively.
Thus, one would expect that the localized
spins reorient themselves to minimize the coupling energy.

 The nonadiabatic term in $H_{A}$ shown as a vector product
of the dot spins that identifies the Dzyaloshinskii-Moriya (DM) interaction,
with the coefficient being the spin current
operator $j_{jk}^{r}$.
It has been demonstrated in ferromagnetic {\it s-d} systems  \cite{Kik16247201,Tat19208}
that the DM interaction induced by current
gives rise of skyrmions \cite{Ros06797, Woo16501}.
Nonadiabatic processes with spin-flipping
arise from the non-diagonal field components,
$A_{jk}^{x}$ and $A_{jk}^{y}$.

 Importantly, it is the adiabatic part that leads to
an effective chiral interaction.
In the adiabatic approximation,
electrons remain in the
spin eigenstates and the flipping between states is forbidden.
The SU(2) gauge field $A_{jk}$ reduces to $A_{jk}^{\rm{ad}}\equiv A_{jk}^{z}$
that is a U(1) gauge field.
While it preserves the spin states, adiabatic process
produces a spin Berry phase.
The inter-site transfer integral would effectively be
$t_{jk}^{\rm{eff}}=t\langle \bm{n}_{j} | \bm{n}_{k}\rangle=te^{i\varphi_{jk}}\cos{\theta_{jk}}$,
where $\theta_{jk}$ is the angle between
two on-site spins,
and $\varphi_{jk}$ is the solid angle spanned by $\bm{n}_{j}$, $\bm{n}_{k}$ and $z$ \cite{Nao06042001,Ye993737}.
Moreover, the polarization energy $U/2$ splits
the spin degeneracy in the rotating frame.
The ground--state spinor field reduces to a simple form,
$\Phi_{j}^{\dagger}\equiv \hat{d}_{j\uparrow}^{\dagger}$,
and similarly for the spin current $j_{jk}=-it(\Phi_{j}^{\dagger}\Phi_{k}-\Phi_{k}^{\dagger}\Phi_{j})$.
For simplicity, consider the continous situation
in which the difference between the direction of two spins is small.
For a steady state, $\nabla \bm{j}=0$,
the current can be expressed as $\bm{j}=\nabla \times \bm{f}$,
with the vector field $\bm{f}=f\bm{\hat z}$
perpendicular to the planar direction.
Integrating by parts, we get
$\bm{j}\cdot \bm{A}^{\rm{ad}}\sim -\bm{f}\cdot \nabla \times \bm{A}^{\rm{ad}}$.
We see that the current source $\bm{f}$ couples
to the curvature of the adiabatic gauge field,
which acts as an effective magnetic field,
\be\label{EFD}
 B^{\rm eff} \equiv (\nabla \times \bm{A}^{\rm{ad}})_{z}
=-\frac{\hbar}{4}\bm{S}\cdot(\nabla_{x}\bm{S} \times \nabla_{y}\bm{S}).
\ee
By given $\bm{S}_{j}\sim \bm{S}_{k}+(\bm{\delta}_{jk}\cdot \nabla) \bm{S}_{k}$,
we replace the differentials by $\bm{S}_{j}$ and obtain the discrete form,
which is the chiral interaction,
$\bm{S}_{1}(\bm{S}_{2}\times \bm{S}_{3})$, among the triple dots.

Based on above derivations, one can see that the coupling energy $H_{A}$
is zero for isolated TTQDs due to the $C_{3v}$ symmetry.
The open TTQD system with finite bias breaks the inversion symmetry
with a bond current that minimizes $H_{A}$.
This indicates a charge current associated with the scalar chirality $\bm{S}_{1}(\bm{S}_{2}\times \bm{S}_{3})$ of the three spins,
which is the solid angle spanned by the three spins,
as schematically shown in \Fig{fig1}(a).

{\it Topological blockade.}
A blockade behavior of transport current has been implied in \Fig{fig1}.
Now, let us demonstrated this effect more concretely.
Note that each dot preserves the local spin with $\epsilon=-U/2=-0.5\,$mV,
and the chiral ground states are
degenerate at $V=0.15$\,meV.
Based on the displayed $I_c$, $I_t$
and $\rho_{q_{\pm}q_{\pm}}$ in \Fig{fig1}(e) and (f),
it is natural to divide the bias range
into three regions: (a) blockade region ($0\le V\le0.15$\,mV),
(b)  coexistence region ($0.15< V\le0.4$\,mV), and (c)
conduction region ($0.4< V\le0.7$\,mV).
Exemplified in \Fig{fig2}(a)--(c)
are the nonequilibrium spectral functions, $\{A_{j}(\omega)\}$,
in these three regions, respectively.

As shown in \Fig{fig2}(a), $A_{j}(\omega)$ of each QD
has a peak near the Fermi level ($\varepsilon_F=0$).
Intuitively, resonance transport current $I_t$
should be introduced even with a small bias value.
However, $I_t\sim 0$ when $V \le 0.15$ mV, as shown in \Fig{fig1}(e).
On the contrary, $I_c$ has finished its semi-period
in that region with the maximum $|I_c|\sim 4.3$ nA.
The observed transport current $I_t\sim 0$ but a significant $I_c$
is a kind of topological blockade, as it is
due to the formation of topological chiral state [cf.~\Fig{fig1}(f)].
The resulted localization of chiral states
does not contribute to the transport current $I_t$.
Let $\gamma \equiv \int_\text{\tiny $V_1$}^\text{\tiny $V_2$}\! |I_{c}|dV
\big/\int_\text{\tiny $V_1$}^\text{\tiny $V_2$}\! I_{t}dV$
be the measure of the topological blockade effect, within
the specific bias zone. Not surprisingly,
this parameter is large
($\gamma = 25$) in the blockade region
of $0\le V\le0.15$\,mV.

Increasing the bias to  $0.15< V\le0.4$ mV, $I_c$
experiences its second semi-period
with a smaller maximum $|I_c|\sim 2.5$ nA,
as shown in \Fig{fig1}(e).
Meanwhile, a noticeable  $I_t$ emerges and increases with $V$.
By referring \Fig{fig2}(b), one can see that $I_t$ is induced
by the excitation at $\omega\sim\pm0.25$\,meV.
This new peak of $A(\omega)$,
which does not existing in the blockade region, grows with $I_t$,
but the chiral state remains localized.
The channel near $\varepsilon_F$ is still blocked
and contributes little to $I_t$.
We refer the present bias zone the coexistence region, where
$I_c$ and $I_t$ simultaneously appear,
with a deceased value of topological blockade parameter ($\gamma = 4$).

 The observed interplay between $I_t$ and $I_c$
can be understood as follows.
Once an electron serves as a carrier of $I_t$
for its transfering from L to R reservoir,
it no longer contributes
to the circular current $I_c$.
Physically, this is related to
the decrease of occupations on
the chiral states as $I_t$ increases;
see \Fig{fig1}(e) versus \Fig{fig1}(f).
In particular, we refer $0.4< V\le0.7$ mV
the conduction region.
The corresponding spectral functions, $\{A_j(\omega)\}$,
are shown in  \Fig{fig2}(c).
The transport excitation peak at $\omega\sim\pm0.25$ meV
grows high enough to produce much large $I_t$ (in the order of nA).
Although both left and right chiral states still exist
near $\varepsilon_F$,
they are almost degenerate and marginally occupied,
resulting in a small value of $I_c$ (in the order of pA).
The topological blockade is lift,
with $\gamma=0.04$ in this region.

{\it Magnetoelectric (ME) effect.}
The ME phenomenon refers to either
the electric--field induced magnetization
or magnetic--field induced electric polarization \cite{Lan13,Dzy60628,%
Sur941259,Kat05057205}.
The differential ME susceptibility is defined through
$\alpha=\mu_{0}\frac{\partial{M}}{\partial{E}}
=\mu_{0}\frac{\partial{I_{c}}}{\partial{V}}$.
We found that the present TTQD in study
acquires $\alpha$ a value in the order of
$1\,\rm{ns\cdot m^{-1}}$ at low temperature \cite{SM3}.
This is two--order of magnitude larger than that in a typical ME material $\rm{Cr_{2}O_{3}}$ \cite{Mun612589,Rod621126}.

 Last but not least, we would like to
elaborate the importance of Coulomb interaction.
For an  open quantum metal ring without electron-electron correlation ($U=0$),  the circling current can induce a finite magnetic moment \cite{Cin10165202}. However, that current is always accompanied by the transport current at small bias, thus it is unmeasurable in the linear-response regime.
For TTQDs with finite $U$, our results show that the Coulomb blockade and topological blockade suppress the transport current and makes the chiral current measurable and  controllable. The chiral current produces a magnetic moment perpendicular to the plane of the triangle.
The magnetic moment is $M=I_{c}a$, where $a$ is the area of the ring.
In a typical QD device, the characteristic length is nanoscale.
Taking $a=100\,\rm{nm}^{2}$ as an example,
then the estimated magnetic moment is the order of  $10^{-25}\rm{A}\cdot \rm{m}^{2}$. The generated magnetic field at the center of the ring is the order of $10^{-8}\rm{T}$, which can be directly observed in experiments by use of a SQUID detector.

In summary, we have demonstrated
a bias-induced chiral current without magnetic field
involved in a triple triangular quantum dot (TTQD)
structure, with both analytical
elaborations and numerical calculations.
The break of inversion symmetry, which lifts the chiral degeneracy,
results from the coupling between adiabatic spin gauge field
and spin current.  The chiral current oscillates
with bias within the Coulomb blockade regime,
indicating that it is possible to control the chiral spin qubit
by purely electrical manipulations.
The localization of chiral states accounts for
the transport current blockade,
which originates from its topological nature.
We also predict that the magnetoelectric susceptibility
of TTQD systems could be two--orders of magnitude
larger than that of conventional magnetoelectric materials.
The bias-induced chiral current may lead to innovative
applications of TTQDs, ranging from magnetoelectric devices
to chiral quantum computation.

The support from the Natural Science Foundation of China
(Grant Nos.~11774418, 11374363,
11674317, 11974348, 11834014 
and 21633006), 
the National Key R\&D Program of China (Grant No.~2018FYA0305800),
and the Strategic Priority Research Program of CAS (Grant No.~XDB28000000)
is gratefully appreciated.

\newpage
\newpage
\onecolumngrid
\vskip5cm
\begin{center}
 {\large{\bf{Supplementary Material for\\ ``Bias-induced chiral current and topological blockade in triple triangular quantum dots''}}}
\end{center}

\section{Hierarchical equations of motion (HEOM) approach}
Hierarchical equations of motion (HEOM) investigates the properties of QDs in both equilibrium and nonequilibrium states via the reduced density operator\cite{doi:10.1063/1.2938087, doi:10.1063/1.3123526,doi:10.1063/1.3602466, PhysRevLett.109.266403,doi:10.1002/wcms.1269}. At the time $t$, the reduced system density operator, $\rho(t)=\rm{tr}_{env}\rho_{T}(t)$, is related to the initial value at time $t_{0}$ via the reduced Liouville-space propagator $\mathcal{G}(t,t_{0})$ with
\begin{equation}
\rho(t)=\mathcal{G}(t,t_{0})\rho(t_{0}).
\end{equation}
Let $\{\ket{\psi}\}$ be an arbitrary basis set defined in the system space, and $\bm{\psi}=(\psi,\psi^\prime)$. Therefore $\rho(\bm{\psi},t)=\rho(\psi,\psi^\prime,t)$. From the Feynman–Vernon influence functional \cite{feynman2000theory}, the path-integral expression for the reduced Liouville-space propagator is 
\begin{equation}
\mathcal{G}(\bm{\psi},t;\bm{\psi}_{0},t_{0})=\int_{\bm{\psi_{0}[t_{0}]}}^{\bm{\psi} [t]}\mathcal{D}\bm{\psi}e^{iS[\psi]}\mathcal{F}[\bm{\psi}]e^{-iS[\psi^\prime]}.
\end{equation}
Here, $S[\psi]$ is the classical action of the reduced system. $\mathcal{F}[\bm{\psi}]$ is the influence functional determined by the Grassmann variables of the system-environment coupling $f_{\alpha i s}^{\dagger}(t)d_{is}[\psi_{s}]+\rm{H.c.}$ The operators $f_{\alpha i s}^{\dagger}(t)$ and $f_{\alpha is}(t)$ are the reservoir operators defined by
\begin{equation}
f_{\alpha i s}^{\dagger}(t)\equiv e^{iH_{\alpha}t}\left(\sum_{k\in \alpha}V^{*}_{\alpha k is}c_{\alpha ks}^{\dagger}\right)e^{-iH_{\alpha}t},
\end{equation}
\begin{equation}
f_{\alpha is}(t)\equiv e^{iH_{\alpha}t}\left(\sum_{k\in \alpha}V_{\alpha k is}c_{\alpha ks}\right)e^{-iH_{\alpha}t}.
\end{equation}
The influence functional can be evaluated using the Wick theorem and the second-order cumulant expansion method, because all the other higher order cumulants are zero at the thermodynamic Gaussian average for non-interaction leads. As a result, the ensemble average of the second-order cumulants are connected to the reservoir correlation functions $C_{\alpha ijs}^{\pm}(t)$, defined as
\begin{equation}\label{lcf1}
C_{\alpha ijs}^{+}(t)=\left\langle f_{\alpha is}^{\dagger}(t)f_{\alpha js}(0) \right\rangle_{\rm{res}},
\end{equation}
\begin{equation}\label{lcf2}
C_{\alpha ijs}^{-}(t)=\left\langle f_{\alpha is}(t)f_{\alpha js}^{\dagger}(0) \right\rangle_{\rm{res}}.
\end{equation}
In which $\langle ... \rangle_{\rm{res}}$ stands for the ensemble average of the reservoirs, and the time translation invariance is used.  All LCFs in other form different from that in \Eq{lcf1} and \Eq{lcf2} are zero because the lead operators $f_{\alpha is}^{\dagger}(t)$ and $f_{\alpha is}(t)$ satisfy Gaussian statistics. The reservoir spectral density function is defined as
\begin{equation}
J_{\alpha ijs}(\omega)\equiv \frac{1}{2\pi}\int_{-\infty}^{\infty}dt e^{i\omega t}\langle \{  f_{\alpha is}(t), f_{\alpha js}^{\dagger}(0) \} \rangle .
\end{equation} 
With the simplified notation $\sigma = +, -$ and $\bar{\sigma}=-\sigma$, the reservoir correlation functions are associated with the spectral density functions via  fluctuation-dissipation theorem
\begin{equation}
C_{\alpha ijs}^{\sigma}(t)=\int_{-\infty}^{\infty}d\omega e^{i\sigma \omega t}f_{\alpha }^{\sigma}(\omega)J_{\alpha ijs}^{\sigma}(\omega).
\end{equation}
In which $J_{\alpha ijs}^{-}(\omega)=J_{\alpha ijs}(\omega), J_{\alpha ijs}^{+}(\omega)=J_{\alpha jis}(\omega)$, and $f_{\alpha }^{\sigma}(\omega)=1/(1+e^{\sigma \beta_{\alpha}(\omega -\mu_{\alpha})})$ is the Fermi-Dirac function for the electron ($\sigma = +$) or hole ($\sigma = -$) at the temperature $\beta_{\alpha}=1/k_{B}T_{\alpha}$. 
For the linear coupling with a non-interacting reservoir, the reservoir spectral density function can be evaluated as $J_{\alpha ijs}(\omega)=\sum_{k}V_{\alpha k is}^{*}V_{\alpha kjs}\delta(\omega-\epsilon_{\alpha k})$. Make use of Wick theorem and Grassmann algebra, the final expression of influence functional $\mathcal{F}$ reads
\begin{equation}
\mathcal{F}[\bm{\psi}]=\exp{\left\lbrace -\int_{t_{0}}^{t}d\tau \mathcal{R}[\tau,\{\bm{\psi}\}] \right\rbrace}.
\end{equation}
In which $\mathcal{R}[\tau,\{\bm{\psi}\}]=\frac{i}{\hbar^{2}}\sum_{\alpha is \sigma}\mathcal{A}_{is}^{\bar{\sigma}}[\bm{\psi}(t)]\mathcal{B}_{\alpha is}^{\sigma}[t,\bm{\psi}]$. Here, $\mathcal{A}_{is}^{\bar{\sigma}}$ and $\mathcal{B}_{\alpha is}^{\sigma}$ are the Grassmann variables defined as
\begin{equation}
 \mathcal{A}_{is}^{\bar{\sigma}}[\bm{\psi}(t)]=d_{is}^{\sigma}[\psi(t)]+d_{is}^{\sigma}[\psi^{\prime}(t)],
\end{equation}
\begin{equation}
 \mathcal{B}_{\alpha is}^{\sigma}[t,\bm{\psi}]=-i\left[ B_{\alpha is}^{\sigma}(t, \psi)-B^{\prime\sigma}_{\alpha i s}(t,\psi^{\prime}) \right],
\end{equation}
with
\begin{equation}
 B_{\alpha is}^{\sigma}(t, \psi)=\sum_{j}\int_{0}^{t}d\tau C_{\alpha i js}^{\sigma}(t-\tau)d_{js}^{\sigma}[\psi(\tau)],
\end{equation}
\begin{equation}
 B_{\alpha is}^{\prime\sigma}(t, \psi^{\prime})=\sum_{j}\int_{0}^{t}d\tau C_{\alpha ijs}^{\bar{\sigma}*}(t-\tau)d_{js}^{\sigma}[\psi^{\prime}(\tau)].
\end{equation}
The LCFs play the role of memory kernels that can be expanded by a series of exponential functions with the implementation of fluctuation-dissipation theorem together with the Cauchy residue theorem and the Pad\'{e} spectrum decomposition scheme of Fermi function \cite{doi:10.1063/1.3602466}
\begin{equation}
 C_{\alpha i js}^{\sigma}(t)=\sum_{m=1}^{M}\eta_{\alpha ijs m}^{\sigma}e^{-\gamma_{\alpha ijsm}^{\sigma}t}.
\end{equation}
Then the bath influence enters the EOMs with M exponentiations. The auxiliary density operators (ADOs) $\left\lbrace \rho_{\bm{j}}^{n}=\rho_{j_{1}\ldots j_{n}} \right\rbrace$ are determined by the time derivative on influence functional. The final form can be cast into a compact form as follows
\begin{equation}
  \begin{aligned}
   \dot\rho^{(n)}_{j_1\cdots j_n} =& -\Big(i{\cal L} + \sum_{r=1}^n \gamma_{j_r}\Big)\rho^{(n)}_{j_1\cdots j_n}
     -i \sum_{j}\!
     {\cal A}_{\bar j}\, \rho^{(n+1)}_{j_1\cdots j_nj}\nonumber \\
   & -i \sum_{r=1}^{n}(-)^{n-r}\, {\cal C}_{j_r}\,
     \rho^{(n-1)}_{j_1\cdots j_{r-1}j_{r+1}\cdots j_n},
\end{aligned}
\end{equation}
where the index $j \equiv (\sigma s n)$ corresponds to the transfer of an electron to or from ($\sigma=+/-$) the impurity state, and the Grassmannian superoperators  ${\cal A}_{\bar j}\equiv {\cal A}_{is}^{\bar \sigma}$ and ${\cal C}_{j}\equiv {\cal C}_{ijs m}^{\sigma}$ are defined via their fermionic actions on an operator $\hat{O}$ as ${\cal A}_{j}\hat{O}\equiv [\hat{d}_{is}^{\hat{\sigma}},\hat{O}]$ and ${\cal C}_{j}\hat{O}\equiv \eta_{j}\hat{d}_{is}^{\sigma}\hat{O}+\eta_{j}^{*}\hat{O}\hat{d}_{is}^{\sigma}$ respectively. The on-dot electron interactions are contained in the Liouvillian of impurities, ${\cal L}\cdot \equiv [H_{\rm{dot}},\cdot]$. Here, $\rho_{0}(t)=\rho(t)={\rm{tr}}_{\rm{res}}\rho_{\rm{total}}(t)$ is the reduced density matrix and $\{\rho_{j_{1}...j_{n}}(t)^{n};n=1,...,L\}$ are auxiliary density matrices with $L$ denoting the truncation level. Usually a relatively low $L$ (say， $L=4$ or $5$) is often sufficient to yield quantitatively converged results. 
 The transient current
  through the electrode $\alpha$ is determined exclusively by the first-tier auxiliary density operators
  \begin{equation}\label{Current}
   I_{\alpha}(t)=e\frac{i}{\hbar^{2}}\sum_{i\mu}tr_{s}\{\rho_{\alpha\mu}^{\dagger}(t)\hat{d}_{is}-\hat{d}_{is}^{\dagger}\rho_{\alpha\mu}^{-}(t)\}.
  \end{equation}
The retarded singl-electron Green's function $G^{r}_{AB}(t)\equiv -i\theta(t)\left< \left\{ \hat A(t),\hat B(0) \right\} \right>$ can be calculated by use of the HEOM-space linear response theory\cite{PhysRevLett.109.266403, doi:10.1002/wcms.1269}.
The spectral function $A_{is}(\omega)$ can be evaluated by taking $\hat A=\hat d_{is}$ and $\hat B=\hat d_{is}^{\dagger}$
\begin{equation}
 A_{AB}(\omega)\equiv \frac{1}{2\pi}\int dt e^{i\omega t}\left< \left\{ \hat A(t),\hat B(0) \right\} \right> = -\frac{1}{\pi}{\rm{Im}}G^{r}_{AB}(\omega).
\end{equation}
\section{derivation of the chiral term}
Start from the isolated TQD Hamiltonian, we perturbatively derive the chiral interaction. The Hamiltonian of an isolated TTQD is
\begin{equation}
H=\sum_{j=1}^{3}\sum_{s}\epsilon_{js}\hat n_{js}+\sum_{j=1}^{3}U_{j}\hat n_{j\uparrow}\hat n_{j\downarrow}+\sum_{j,k=1}^{3}\sum_{s}(\tilde{t}_{jk}\hat d_{js}^{\dagger}\hat d_{k+1s}+\rm{H.c.}).
\end{equation}
For symmetric gauge we choose $\tilde{t}_{jk}=e^{i\frac{2\pi}{3}\phi/\phi_{0}}t_{jk}$ for $j<k$ and $\tilde{t}_{jk}=e^{-i\frac{2\pi}{3}\phi/\phi_{0}}t_{jk}$ for $j>k$. Separate the kinetic part of the Hamiltonian into three terms $T_{0}$, $T_{1}$ and $T_{-1}=T_{1}^{\dagger}$, $H=H_{0}+V+T_{1}+T_{-1}+T_{0}$. Where $T_{m}$ changes the number of doubly occupied sites by $m$ when it acts on a state
\begin{equation}\label{tx}
T_{0}=\sum_{j,k=1}^{3}\sum_{s}\tilde{t}_{jk}\left[\hat n_{j\bar{s}}\hat d_{js}^{\dagger}\hat d_{ks}n_{k\bar{s}}+(1-\hat n_{j\bar{s}})\hat d_{js}^{\dagger}\hat d_{ks}(1-\hat n_{k\bar{s}})\right],
\end{equation}
\begin{equation}\label{ty}
T_{1}=\sum_{jks}\tilde{t}_{jk}\hat n_{j\bar{s}}\hat d_{js}^{\dagger}\hat d_{ks}(1-\hat n_{k\bar{s}}),
\end{equation}
\begin{equation}\label{tz}
T_{-1}=\sum_{jks}\tilde{t}_{jk} (1-\hat n_{j\bar{s}})\hat d_{js}^{\dagger}\hat d_{ks}\hat n_{k\bar{s}}.
\end{equation}
Denote $V=\sum_{j}U_{j}\hat n_{j\uparrow}\hat n_{j\downarrow}$, $H_{0}=\sum_{js}\epsilon_{js}\hat n_{js}$.
At half-filling, We use a unitary transformation to restrict the Hilbert space into the singly occupied subspace
\begin{equation}
H^{\prime}=e^{S}He^{-S} = H+[S,H]+\frac{1}{2}[S,[S,H]]+\frac{1}{6}[S,[S,[S,H]]].
\end{equation}
In which the unitary matrix $S=\frac{1}{U}(T_{1}-T_{-1})+\frac{1}{U^{2}}([T_{1},T_{0}]+[T_{-1},T_{0}])$ eliminates the hopping between states with different numbers of doubly occupied sites

\begin{equation}
H^{\prime}=H_{0}+V+T_{0}+\frac{1}{U}[T_{1},T_{-1}]+\frac{1}{U^{2}}\left[ T_{1}T_{0}T_{-1}+T_{-1}T_{0}T_{1}-\frac{1}{2}(T_{1}T_{-1}T_{0}+T_{-1}T_{1}T_{0}+T_{0}T_{1}T_{-1}+T_{0}T_{-1}T_{1}) \right].
\end{equation}
However, the transformed Hamiltonian is still defined in the whole Hilbert space. A low energy projection is needed to get the effective Hamiltonian restricted in the subspace that without any doubly occupied state. Introduce the operator $P=\prod_{i}(1-n_{i\uparrow}n_{i\downarrow})$ that project the system into the subspace that the doubly occupied states are not included. The operator $Q=1-P$ contains all possible doubly occupied states. Notice $PVP=0$, $PT_{m}...T_{n}P=0$ if $m\neq -1$ and $n\neq 1$, we get
\begin{equation}\label{heff}
 PH^{\prime}P=H_{0}-\frac{1}{U}T_{-1}T_{1}+\frac{1}{U^{2}}T_{-1}T_{0}T_{1}.
\end{equation}
Substitute \Eq{tx}, \Eq{ty} and \Eq{tz} into \Eq{heff}, the operation of the annihilation and creation operator is straightforward. We get
\begin{equation}
PH^{\prime}P=\sum_{j=1}^{3}\epsilon_{j}\hat n_{j}+\sum_{j,k=1}^{3}J_{jk}(\hat{\bm{S}}_{j}\cdot \hat{\bm{S}}_{k}-\frac{1}{4}\hat n_{j}\hat n_{k})+\chi\hat{\bm{S}}_{1}\cdot (\hat{\bm{S}}_{2}\times \hat{\bm{S}}_{3}).
\end{equation}
In which $J_{jk}=4t^{2}/U$, and $\chi=24t^{3}/U^{2}\sin(2\pi\phi/\phi_{0})$. For half-filling $n=1$, we get
\begin{equation}
PH^{\prime}P=\sum_{j=1}^{3}\epsilon_{j}+\sum_{j,k=1}^{3}J_{jk}(\hat{\bm{S}}_{j}\cdot \hat{\bm{S}}_{k}-\frac{1}{4})+\chi\hat{\bm{S}}_{1}\cdot (\hat{\bm{S}}_{2}\times \hat{\bm{S}}_{3}).
\end{equation}

\section{spin gauge field, berry phase and chiral interaction}
The action of the TTQD Anderson impurity reads $S_{0}+S_{\rm{int}}$,
\begin{equation}\begin{split}
S_{0}[\psi^{*},\psi]=\int_{0}^{\beta}d\tau \bigg[ \sum_{j=1}^{3}\sum_{s}\psi_{js}^{*}(\partial_{\tau}-\mu)\psi_{js}-t\sum_{j,k=1}^{3}\sum_{s}(\psi_{js}^{*}\psi_{ks}+\rm{c.c.}) \bigg]
\end{split},\end{equation}
\begin{equation}
S_{\rm{int}}[\psi^{*},\psi]=U\int_{0}^{\beta}d\tau \sum_{j=1}^{3}\psi_{j\uparrow}^{*}\psi_{j\uparrow}^{\uparrow}\psi_{j\downarrow}\psi_{j\downarrow},
\end{equation}
in which $\psi$ is a Grassmannian field. To describe the charge and spin fluctuations, we write the interaction action
\begin{equation}
 S_{\rm{int}}[\psi^{*},\psi]=\frac{U}{4}\int_{0}^{\beta}d\tau \sum_{j=1}^{3}\left[(\Psi_{j}^{\dagger}\Psi_{j})^{2}-(\Psi_{j}^{\dagger}\sigma^{z}\Psi_{j})^{2}\right],
\end{equation}
where $\Psi_{j}^{\dagger}=(\hat d_{j\uparrow}^{\dagger},\hat d_{j\downarrow}^{\dagger})$ is a two-component spinor. By introducing two real auxiliary fields, we rewrite the partition function
\begin{equation}\begin{split}
 Z=\int \mathcal{D}[\Delta_{c},\Delta_{s},\Psi^{\dagger},\Psi]\exp\left\{ -S_{0}[\Psi^{\dagger},\Psi]-S_{\rm{int}}[\Psi^{\dagger},\Psi]\right\}\label{ptf}
\end{split},\end{equation}
with the interaction action
\begin{equation}
 S_{\rm{int}}=\int_{0}^{\beta}d\tau \sum_{j}  \left[ \frac{1}{U}(\Delta_{cj}^{2}+\Delta_{sj}^{2})-\Psi_{j}^{\dagger}(i\Delta_{cj}+\Delta_{sj}\sigma^{z})\Psi_{j}\right].
\end{equation}
The decoupled action breaks at least formally the spin-rotational invariance of \Eq{ptf}, which does not reproduce the Hatree-Fock approximation at the saddle-point level.
Because the choice of spin quantization axis is arbitrary, we can restore the invariance of the action by rotating the quantization axis $\sigma^{z}\rightarrow \bm{\sigma}\cdot \bm{\Omega}$ and performing an angular integration over a site- and time-dependent unit vector $\bm{\Omega}$. The the interaction action reads
\begin{equation}
 Z=\int \mathcal{D}[\bm{\Omega}]Z[\bm{\Omega}],
\end{equation}
with the interaction action
\begin{equation}\begin{split}
 S_{\rm{int}}=\int_{0}^{\beta}d\tau \sum_{j} \bigg[ \frac{1}{U}(\Delta_{cj}^{2}+\Delta_{sj}^{2})-\Psi_{j}^{\dagger}(i\Delta_{cj}+\Delta_{sj}\bm{\sigma}\cdot \bm{\Omega})\Psi_{j}\bigg]
\end{split}.\end{equation}
Perform a unitary transformation on the Grassmann field $\Phi_{j}=R_{j}\Psi_{j}$, where $R_{j}$ is a site- and time-dependent $SU(2)$ rotation matrix satisfying $\bm{\sigma}\cdot \bm{\Omega}_{j} =R_{j}\sigma^{z}R_{j}^{\dagger}$. 
We parameterize $\bm{\Omega}_{j}=(\sin{\theta_{j}}\cos{\phi_{j}},\sin{\theta_{j}}\sin{\phi_{j}},\cos{\theta_{j}})$ and the action takes the form of
\begin{equation}
 S[\Phi^{\dagger},\Phi]=S_{\rm{sp}}+S_{1}+S_{2},
\end{equation}
where 
\begin{equation}
S_{\rm{sp}}=\int_{0}^{\beta}d\tau \sum_{j} \bigg[ \frac{1}{U}(\Delta_{cj}^{2}+\Delta_{sj}^{2})-\Phi_{j}^{\dagger}(\partial_{\tau}-\mu-i\Delta_{cj}+\Delta_{sj}\sigma^{z})\Phi_{j}\bigg]
\end{equation}
being the saddle-point action and 
\begin{equation}
S_{1}=\int_{0}^{\beta}d\tau \sum_{j}\Phi_{j}^{\dagger}R_{j}^{\dagger}\partial_{\tau}R_{j}\Phi_{j}, \quad S_{2}=-\int_{0}^{\beta}d\tau \sum_{jk} t\left[\Phi_{j}^{\dagger}(R_{k}^{\dagger}R_{j}-1)\Phi_{k}+\rm{c.c.}\right]
\end{equation}
being the coupling action. 
Taking uniform and time-independent auxiliary fields $\Delta_{cj}=\Delta_{c}, \Delta_{sj}=\Delta_{s}$, the saddle-point values $\Delta_{c}$ and $\Delta_{s}$ are obtained from saddle-point equations $\partial_{\Delta_{c}}\ln{Z_{\rm{sp}}}=0$ and $\partial_{\Delta_{s}}\ln{Z_{\rm{sp}}}=0$,
\begin{equation}
 \frac{2}{U}\Delta_{c}=i\langle \Psi_{j}^{\dagger}\Psi_{j}\rangle=n, \quad \frac{2}{U}\Delta_{s}=i\langle \Psi_{j}^{\dagger}\sigma^{z}\Psi_{j}\rangle=m .
\end{equation}
The $SU(2)$ gauge field is defined by the unitary transformation $R$
\begin{equation}
 A_{0j}=-iR_{j}^{\dagger}\partial_{\tau}R_{j},\quad A_{jk}=-i(R_{j}^{\dagger}R_{k}-1),
\end{equation}
which can be expressed by use of Pauli matrices
\begin{equation}
 A_{0j}=A_{0j}^{r}\sigma_{r}=\bm{A}_{0j}\cdot \bm{\sigma},\quad A_{jk}=A_{jk}^{r}\sigma_{r}=\bm{A}_{jk}\cdot \bm{\sigma},
\end{equation}
where $r=x,y,z$ is the direction in spin space. We introduce the spin density and spin current fields in the rotated frame
\begin{equation}
 \tilde{s}_{j}^{r}=\Phi^{\dagger}_{j}\sigma_{r}\Phi_{j}, \quad \tilde{j}_{jk}^{r}=-it(\Phi_{j}^{\dagger}\sigma_{r}\Phi_{k}-\Phi_{j}^{\dagger}\sigma_{r}\Phi_{j}).
\end{equation}
Retain the first order cumulants
\begin{equation}
 S[\Phi^{\dagger},\Phi]=\langle S_{1} + S_{2} \rangle_{\rm{sp}},
\end{equation}
where $\langle \cdots \rangle_{\rm{sp}}$ are to be calculated with the saddle-point action $S_{\rm{sp}}$.  The first-order cumulants $\langle S_{1}\rangle$ and $\langle S_{2}\rangle$ are given by
\begin{equation}
S_{1}=i\int_{0}^{\beta}d\tau \sum_{j} \langle \tilde{s}_{j}^{r}\rangle A_{0j}^{r}, \quad S_{2}=\int_{0}^{\beta}d\tau \sum_{jk} \langle \tilde{j}_{jk}^{r} \rangle A_{jk}^{r}.
\end{equation}
The spin density and spin current operators are related to which in the laboratory frame as $\tilde{j}_{jk}^{r}=\sum_{q}j_{jk}^{q}\gamma_{qr}$, where $\gamma_{qr}=2m_{r}m_{q}-\delta^{rq}$ is the $SO(3)$ rotation matrix corresponding to $R$, with $\bm{m}_{j}=[\sin{\theta_{j}/2}\cos{\phi_{j}}, \sin{\theta_{j}/2}\sin{\phi_{j}}, \cos{\theta_{j}/2}]$. Then we obtain the coupling action in laboratory frame,
\begin{equation}
S_{1}=i\int_{0}^{\beta}d\tau \sum_{j} \langle s_{j}^{r}\rangle A_{0j}^{r}, \quad S_{2}=\int_{0}^{\beta}d\tau \sum_{jk} \langle j_{jk}^{r} \rangle A_{jk}^{r}.
\end{equation}
The matrix $R_{j}$ can be explicitly written as 
\begin{equation}
R_{j}=\left[
 \begin{matrix}
   \cos{\frac{\theta_{j}}{2}} & e^{-i\phi_{j}}\sin{\frac{\theta_{j}}{2}} \\
   e^{i\phi_{j}}\sin{\frac{\theta_{j}}{2}} &   -\cos{\frac{\theta_{j}}{2}}
  \end{matrix}\right]=\bm{m}\cdot \bm{\sigma}.
\end{equation}
 Making use of the identity $R^{\dagger}R=1$, $A_{jk}$ can be expressed as $A_{jk}=-iR_{j}^{\dagger}(R_{k}-R_{j})$. Denote $\bm{r}_{j}=\bm{r}+\delta\bm{r}$ We can expand $R$ with respect to position coordinates $R[\bm{r}+\delta\bm{r}]=R[\bm{r}]+\delta\bm{r}\nabla R[\bm{r}]+\mathcal{O}(\delta\bm{r}^{2})$ to obtain $A_{\bm{r},\bm{r}+\delta\bm{r}}=-iR[\bm{r}]^{\dagger}(R[\bm{r}+\delta\bm{r}]-R[\bm{r}])=-iR[\bm{r}]\delta\bm{r}\nabla R[\bm{r}]$. By use of Eq.(14), we get
\begin{equation}
 \bm{A}_{\bm{r},\bm{r}+\delta\bm{r}}=\frac{1}{2}\bm{\Omega}_{r}\times \delta\bm{r}\nabla\bm{\Omega}_{r}-A_{\bm{r},\bm{r}+\delta\bm{r}}^{z}\bm{\Omega}_{r}.
\end{equation}
Where $A_{\bm{r},\bm{r}+\delta\bm{r}}^{z}=(1-\cos{\theta})\nabla \phi \cdot \sigma_{z}$ is the adiabatic spin gauge field (Berry phase). 
The polarization energy $U/2$ splits the spin degeneracy in the rotated frame with the high energy electrons neglected, the spinor field reduced to a simple form $\Phi_{j}^{\dagger}\equiv \hat d_{j\uparrow}^{\dagger}$ and similarly for the spin current $j_{jk}=-it(\Phi_{j}^{\dagger}\Phi_{k}-\Phi_{k}^{\dagger}\Phi_{j})$. For simplicity, considering the continuous situation.
For steady state $\nabla \cdot \bm{j}=0$, and $\bm{j}$ can be expressed in form of the curl of vector field $\bm{j}=\nabla \times \bm{f}$. Integrating by parts, we get $\bm{j}\cdot \bm{A}^{z}\sim - \bm{f}\cdot \nabla \times \bm{A}^{z}$. We see that the curvature acts as a effective magnetic field $(\nabla \times \bm{A}^{z})_{z}=-\frac{\hbar}{4}\bm{S}\cdot(\nabla_{x}\bm{S} \times \nabla_{y}\bm{S})$. By given $\bm{S}_{j}\sim \bm{S}_{k}+(\bm{\delta}_{jk}\cdot \nabla) \bm{S}_{k}$,
we replace the differentials by $\bm{S}_{j}$ and obtain the discrete form,
which is the chiral interaction 
$\bm{S}_{1}(\bm{S}_{2}\times \bm{S}_{3})$ among the triple dots.
\section{Bond current and chiral current}

For isolated TTQD, the chiral current can only be driven by a local magnetic flux, and its magnitude is evaluated by the bond current $I_{jk}$, i.e., the current flowing between the $j$th dot and the $k$th dot
\begin{equation}
 \hat{I}_{jk}=i\frac{e}{\hbar}\sum_{s}(t_{jk}\hat d_{js}^{\dagger}\hat d_{ks}- \rm{H.c.}).
\end{equation}
It is identical to the chiral current  defined by use of chiral operator [cf. Eq.(3) in the main text]. However, for the open TTQD connected with two electrodes, the bond current is not equal to the chiral current because the existence of the transport current destroys its continuity. Because of the locality, the bond currents of steady state satisfy Kirchhoff’s current law $I_{12}+I_{t}=I_{23}$ and $I_{23}+I_{13}=I_{t}$ , where $I_{t}$ is the transport current measured at the electrode for steady state. Thus it is ambiguous whether $I_{12}$ or $I_{23}$ describes the chiral current. This is indeed related to requirement of the gauge invariance of the chiral current \cite{PhysRevB.81.165202}. To eliminate this ambiguity, we have to find a observable that contain the global characteristic of the electron exchange, which is the chiral current defined by use of chiral operator.

\begin{figure}[H]
\centering
\includegraphics [width=3.0in]{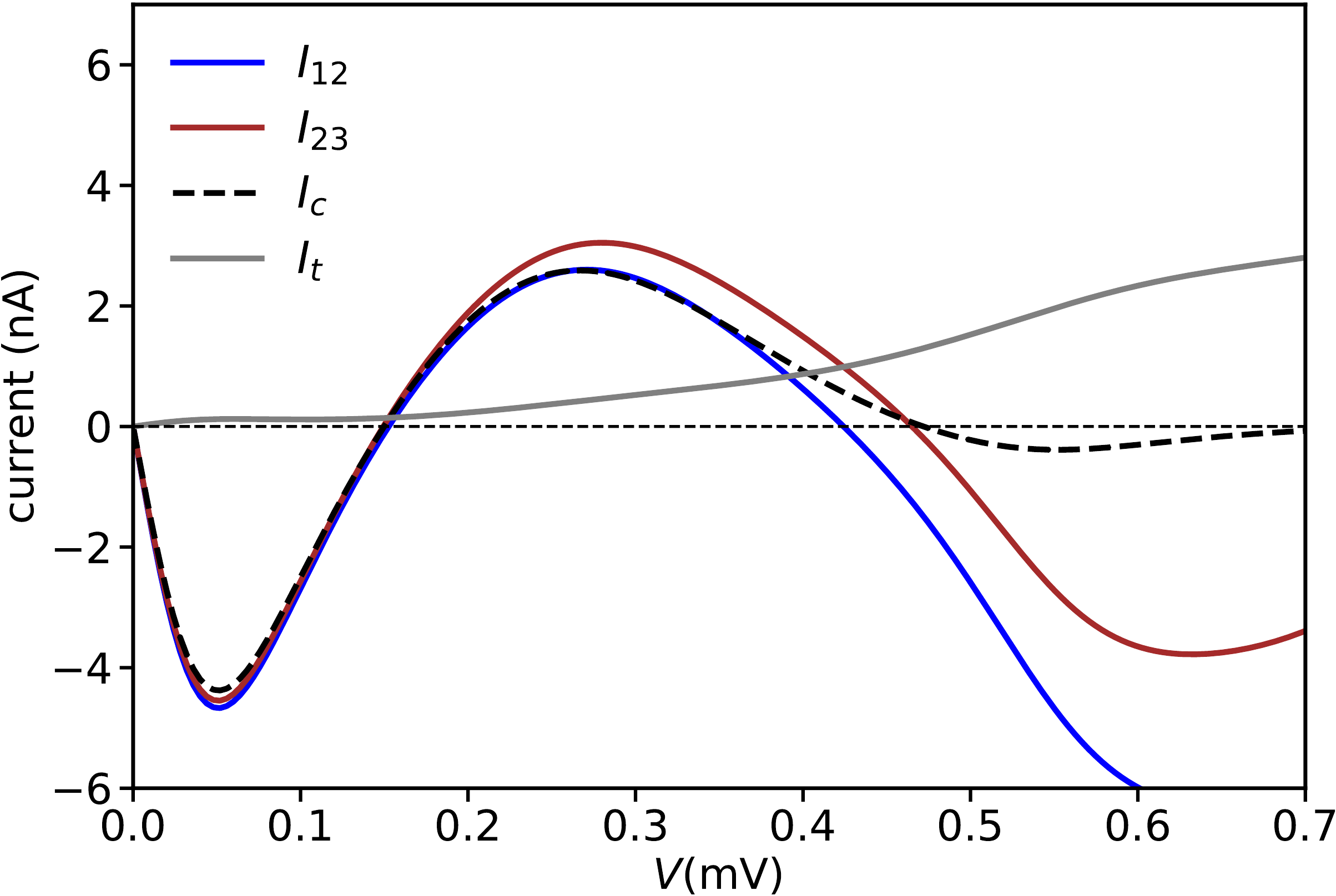}
\caption{(color online). Bond current, chiral current and transport current versus bias. The triangle-up marker and triangle-down marker correspond to the clockwise and anticlockwise vertex current are schematically shown in Fig.1 in the main text. The other parameters are $\Delta=0.025$, $t=0.25$, $T=0.05$, $\epsilon=-0.5$, $U=1.0$.}\label{fig3}
\end{figure}

The comparison between chiral and bond current is shown in \Fig{fig3}. The bond current $I_{13}$ is not depicted because of the Kirchhoff law $I_{13}=-I_{12}$. Under the
restriction of the Kirchhoff’s law, the bond currents are favorable to have a unified magnitude and toroidal direction to minimize the coupling energy $H_{A}$. Therefore the bond current can be approximately regarded as the chiral current in small bias region, suggesting a nearly pure circling current with small leakage. At a higher voltage, the inter-dot current $I_{12}$ and $I_{23}$ splits and transport current increases rapidly, this is because the double occupation energy level and single particle energy level moves towards the Fermi energy of the electrodes. 

\section{Magnetoelectric effect}
\begin{figure}[htb]
\centering
\includegraphics [width=3.0in]{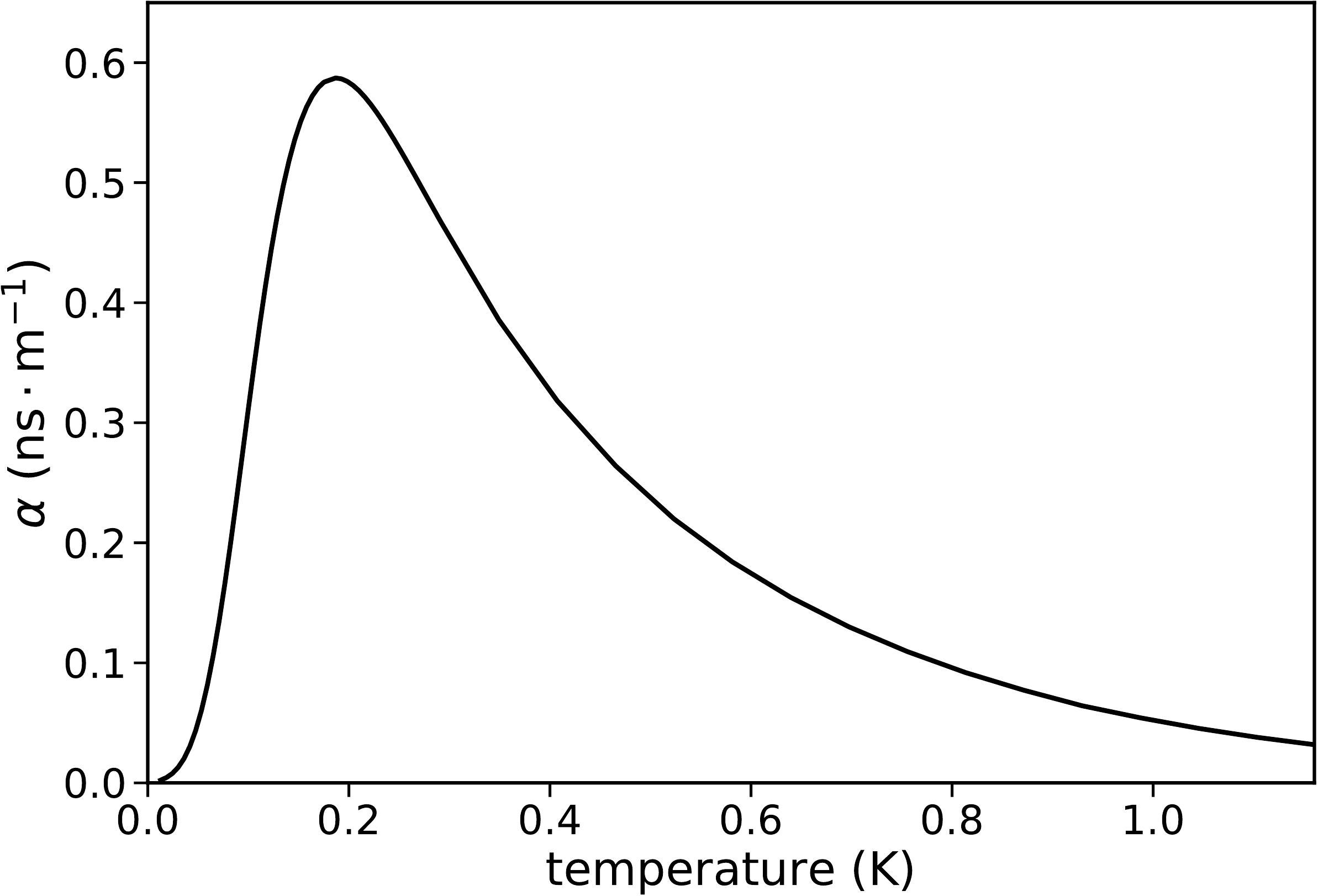}
\caption{Magnetoelectric susceptibility of the TTQD as a function of temperature. The other parameters are the same as those used in \Fig{fig1} in the main text.}\label{fig4}
\end{figure}

\end{document}